# Evaluating ChatGPT's Decimal Skills and Feedback Generation in a Digital Learning Game

Huy A. Nguyen[1] [0000-0002-1227-6173], Hayden Stec[1], Xinying Hou[2] [0000-0002-1182-5839], Sarah Di[1], and Bruce M. McLaren[1]

[1] Carnegie Mellon University, Pittsburgh PA 15213, USA
`hn1@cs.cmu.edu`
[2] University of Michigan, Ann Arbor MI 48104, USA

**Abstract.** While open-ended self-explanations have been shown to promote robust learning in multiple studies, they pose significant challenges to automated grading and feedback in technology-enhanced learning, due to the unconstrained nature of the students' input. Our work investigates whether recent advances in Large Language Models, and in particular ChatGPT, can address this issue. Using decimal exercises and student data from a prior study of the learning game *Decimal Point*, with more than 5,000 open-ended self-explanation responses, we investigate ChatGPT's capability in (1) solving the in-game exercises, (2) determining the correctness of students' answers, and (3) providing meaningful feedback to incorrect answers. Our results showed that ChatGPT can respond well to conceptual questions, but struggled with decimal place values and number line problems. In addition, it was able to accurately assess the correctness of 75% of the students' answers and generated generally high-quality feedback, similar to human instructors. We conclude with a discussion of ChatGPT's strengths and weaknesses and suggest several venues for extending its use cases in digital teaching and learning.

**Keywords:** ChatGPT · Decimal Numbers · Automated grading · Self-Explanation

## 1 Introduction

As Large Language Models (LLMs) have become more accessible in recent years, questions have risen across scientific domains about the potential applications of these powerful tools. Within the field of education, there is noticeable excitement about the possibility of incorporating LLMs into digital learning platforms [14, 18, 23, 28]. Past research in the area of assessment and evaluation has yielded promising results from LLMs [20, 28], and with the availability of more powerful and generalizable models like ChatGPT [32], it is now feasible to ask whether these technologies can go beyond traditional classification tasks to generating high-quality, human-like feedback that benefit both students and teachers alike.

Our work explores one such use case in the context of *Decimal Point*, a game that teaches decimal numbers and operations to middle school students. *Decimal Point* features a series of mini-games, where the player solves a decimal problem (e.g., placing 0.456 on a number line between -1 and 1) and then responds to a self-explanation prompt (e.g., "Is 0.456 to the left or right of the number line? How do you know?"). The game has been deployed in many classroom studies to explore a variety of learning game topics [9, 10, 17], with a recent study showing that open-ended self-explanation prompts yielded the best learning outcomes, compared to multiple-choice and drag-and-drop formats, without disrupting the game play experience [17]. However, grading the open-ended responses was highly labor-intensive and could not be carried out in real-time [21], thus limiting the students' opportunities to learn from timely feedback. This drawback motivates our current work, which explores the capabilities of ChatGPT to serve as a "teaching assistant" tasked with grading and generating feedback for students' self-explanations in the game. In particular, we focused on three research questions:

**RQ1:** Can ChatGPT correctly answer the problem-solving and self-explanation questions in the game *Decimal Point*?

**RQ2**: Can ChatGPT accurately rate the correctness of students' self-explanation answers?

**RQ3**: Can ChatGPT provide *instructionally meaningful* feedback to incorrect self-explanation answers?

Through the investigation of these questions, our work makes the following contributions to the area of technology-enhanced learning. First, we report the evaluation results of a state-of-the-art LLM on a large volume of real student data, with actionable insights into its strengths and weaknesses. Second, we introduce a generalizable workflow for assessing the pedagogical value of ChatGPT's outputs, including a 6-category coding rubric that can be applied to other knowledge domains. Finally, we have open sourced our ChatGPT querying scripts to support reproducibility and assist other researchers in their own explorations of LLMs' potential.

## 2   Background

### 2.1   Self-explanation in Digital Learning Games

Prompted self-explanation has been shown to support meaningful learning and help develop metacognitive skills within digital learning games. For instance, [11] reported that multiple-choice self-explanation prompts within a digital learning environment aimed towards helping students acquire scientific concepts about light and shadow led to higher learning gains than no self-explanation. Additionally, [22] found that within a middle school fractions game, menu-based self-explanation prompts aimed towards helping students draw connections between game terminology and mathematics terminology were more effective for learning than no self-explanation prompts.

However, not all studies support incorporating self-explanation into learning games. [1] reported that in the game *Newtonian Game Dynamics*, there were no learning differences between students in the self-explanation condition with explanatory feedback and the control condition with neither self-explanation nor explanatory feedback. In addition, students in the experimental condition completed fewer game levels than the control group, demonstrating that self-explanation has the potential to disrupt game flow and introduce extraneous cognitive processing. Wylie and Chi [31] posit that this inconsistency in learning differences can be attributed to the *type* of prompted self-explanation used in digital learning games. In particular, there are five main types of self-explanation cast along a continuum from *unconstrained self-* *explanation* prompts, where learners create their own explanations without guidance, to *highly constrained self-explanation* prompts (i.e., *menu-based self-explanations*), where learners choose from a given set of explanations. According to Johnson & Mayer [12], highly constrained self-explanations present less cognitive load to learners and could therefore lead to better learning outcomes. On the other hand, Wylie and Chi [31] argue that less constrained self-explanation prompts foster more *constructive* and *active* engagement with the learning material, whereby the learner is prompted to generate their own explanations (constructive) and type or write them down (active). In turn, these activities enable learners to connect their prior knowledge, fill in gaps of understanding and achieve better learning outcomes.

Other than their cognitive loads, a major difference between these self-explanation formats lies in the ability to provide feedback to students. Highly constrained questions typically correspond to a fixed set of solutions that can be checked by a computer system, thus enabling real-time feedback generation. On the other hand, unconstrained self-explanations are difficult to grade in an automated manner, due to challenges in understanding student inputs in the presence of varied writing styles, grammatical errors and misspellings [25]. Even so, automated grading of open-ended answers remains a crucial area of research, as it can enable timely feedback that helps students improve their learning [33]. More recently, the development of large language models may provide a pathway towards this capability in a wide range of learning platforms. In the following sections, we further describe recent advances in language models and our evaluation of this hypothesis.

## 2.2   Large Language Models in Education

Large language models (LLMs) generally function by leveraging massive corpora of textual data as input to transformer-based models, consisting of hundreds of billions of parameters [7]. The hallmark of the transformer architecture is the attention mechanism, which helps these models make better predictions by focusing on specific parts of the text that are relevant to the task at hand. Among the prominent LLMs is the Generative Pretrained Transformer (GPT) series developed by OpenAI, with the recent introduction of their GPT-3.5 model, released publicly as ChatGPT [32]. This model has been shown to excel at question-answering and various reasoning tasks, while demonstrating superior generalization compared to previous LLMs [24].

The release of ChatGPT has prompted a new wave of research on how it can be best applied in educational contexts [28]. Recent work has shown that ChatGPT can act as a tutor to reduce students' cognitive load in medical education [14], or generate algebra hints that improve student learning [23]. At the same time, there have been concerns about potential cheating or plagiarism in writing assignments, fueled by the ease at which ChatGPT can generate human-like writing within seconds [5]. While much of this debate centers around how students should and should not use ChatGPT, less attention has been paid to whether teachers and instructors can benefit from this technology. For example, [18] has shown that a GPT-3 model is able to help teachers detect high-quality student-generated questions that can be incorporated into future iterations of a college-level chemistry course. Our work seeks to further explore use cases in this direction, especially those that involve reviewing a significant amount of text data and typically require considerable efforts from human experts. In particular, we investigate whether ChatGPT can provide accurate assessments of, and high-quality feedback to, more than 5,000 self-explanation responses produced by students who play a digital learning game for decimal numbers. To our knowledge, this is the first rigorous evaluation of ChatGPT on a large volume of real student data. We elaborate on the data source and our evaluation criteria below.

### 2.3 The Learning Game *Decimal Point*

*Decimal Point* is a single-player, non-competitive learning game targeted at helping middle school students practice decimal numbers and operations [15]. The game specifically targets common decimal misconceptions held by students, such as longer decimals are larger than shorter decimals (e.g., 0.234 > 0.9). In the game, the student travels through a fictional amusement park with alien friends and completes a series of 24 mini-games. Each mini-game features two problem-solving questions, followed by a self-explanation prompt designed to address common decimal misconceptions and reinforce student learning (Figure 1). There are five types of problem-solving questions; each type corresponds to specific templates for the self-explanation prompt that follows (Table 1). Throughout their game play, the student gets immediate feedback about the correctness of their answers, and can make any number of attempts to arrive at correct answers. In addition, they must answer all questions correctly to advance to the next mini-game round.

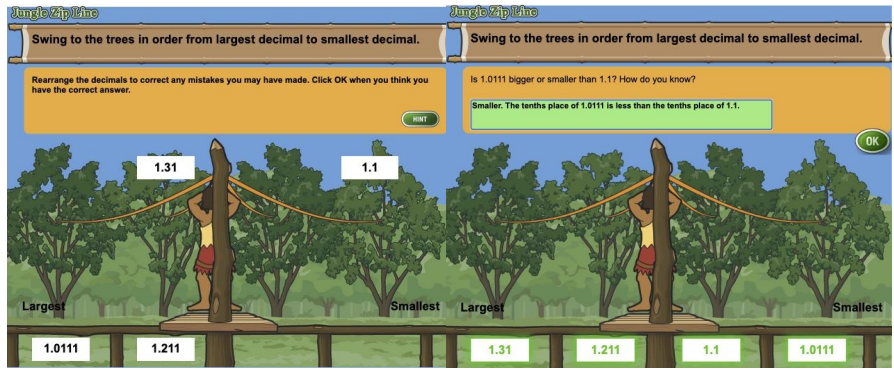

**Fig. 1.** A problem-solving question (left) and self-explanation prompt (right) in the mini-game *Jungle Zipline*. The student sorts a sequence of decimal numbers and then explains how to compare two numbers in this sequence.

**Table 1.** The types of problem-solving activities and the corresponding examples for self-explanation prompts.

| Problem Type | Description | Self-explanation templates |
| --- | --- | --- |
| *Addition* | Enter the carry and result digits when adding two decimal numbers. | What should you do when there are more than 10 in any column? |
| *Bucket* | Label each given number as "bigger than" or "smaller than" a threshold | Is number X bigger or smaller than number Y? How do you know? |
| *Number Line* | Locate the position of a number on a number line between -1 and 1 | Is number X closer to 0 or to -1 / 0.5 / 1? How do you know? |
| *Sequence* | Find the next two terms in an increasing or decreasing arithmetic sequence | What should you remember to find the next number in the sequence? When should you carry / how do you know? |
| *Sorting* | Sort the given decimal numbers in increasing or decreasing order | Is number X bigger or smaller than number Y? How do you know? |

*Decimal Point* has been used in many classroom studies to explore different learning game topics, including, for instance, gender effects [10], open student models [9], and the benefits of self-explanation [17]. While most of these prior studies implemented multiple-choice self-explanation questions, we also experimented with two other formats, drag-and-drop and open-ended, in a study reported in [17]. Our results indicated that open-ended self-explanation prompts led to the best learning outcomes; however, because the students' responses are unstructured text, they could not be graded in real-time. Instead, we implemented a keyword matching heuristic, where the student's answer was considered acceptable if it contained more than three words, with at least one of the words matching a set of keywords that might be found in a correct explanation [17]. In this work, we investigate whether ChatGPT could address this gap by grading the students' answers and providing meaningful instructional feedback in a real-time and automated manner.

### 2.4 Types of Feedback and Evaluation

White the positive effects of feedback in digital learning platforms is well known [30], its benefits can be impacted by a variety of factors, including the timing and content of the feedback [29]. In particular, immediate feedback allows students to quickly identify and correct any mistakes or misconceptions they may have, which is especially helpful for students with low prior knowledge [26]. At the same time, delayed feedback has been shown to benefit learning in complex domains, where the longer intervals enable more time for reflection and knowledge assimilation [4].

Feedback content typically consists of a verification component – i.e., a simple "right" or "wrong" statement – and an elaboration component, which justifies the verification [13]. The elaboration may contain different kinds of information, including (1) task-specific information, such as the correct answer and relevant rubric items, (2) instruction-based information, which explains why an answer is correct or incorrect, and (3) extra-instructional information, which provides further worked examples or analogies not included in the original learning materials. Under this categorization schema, prior research has reported that feedback messages with both the verification and elaboration components can lead to better learning than those that only feature verification [30]. At the same time, the affective nature of the language used in feedback messages may also play a role in its effectiveness – for instance, encouraging and polite language can help students improve their learning, especially those with low prior knowledge [16].

In the context of *Decimal Point*, as the decimal exercises focus on addressing common decimal misconceptions [15], the use of immediate feedback with clear elaborations and positive language would provide the greatest benefit to students. Thus, with the introduction of ChatGPT, we are interested in whether the game can generate feedback messages with the desired qualities. In what follows, we further describe the properties of the decimal exercises in the game and our rubric for assessing ChatGPT's responses.

## 3 Methods and Materials

### 3.1 Dataset

We employed student data from a classroom study conducted by McLaren and colleagues [17] in the spring of 2021. In this study, 357 5th and 6th graders were randomly assigned to play the learning game *Decimal Point* with either multiple-choice, drag-and-drop, or open-ended self-explanation questions. For the purpose of our analysis, we focused on students' open-ended self-explanations and ChatGPT's capability in evaluating them. Thus, our final dataset contains 117 students who generated 5,142 open-ended self-explanation answers in total. Based on the manual coding of these answers, as reported in [21], there were 1,000 correct, 4,076 incorrect, and 66 off-topic answers, which we also treated as incorrect.

### 3.2 ChatGPT Queries

Our data collection with ChatGPT took place in the conversational web interface at https://chat.openai.com/chat, using the March 23 version of ChatGPT[1] and with GPT-3.5 as the underlying model [32]. We also utilized the ChatGPT Wrapper library[2] to automate the process of sending questions to ChatGPT and recording its responses. For RQ1 – whether ChatGPT can correctly answer the problem-solving and self-explanation questions in *Decimal Point* – ChatGPT's answers were evaluated by a researcher in the team who is experienced with middle school math education.

For RQ2 – whether ChatGPT can correctly label students' open-ended self-explanations as correct or incorrect – we made use of the student answers from [17] and the rubric for grading them from [21]. In particular, we constructed a query template (Figure 2) that provides ChatGPT with the following information:

- The self-explanation prompt (se_prompt), for example: "When adding two decimal numbers, how should you line up the numbers and add them?"
- The student's answer (student_answer), for example: "On top of the same values."
- The relevant grading rubric items (rubric_items) for this prompt, for example: "A correct response should mention aligning numbers by either the decimal point or decimal place values."

> Given a student's self-explanation response after they've played an educational game about decimal numbers, label each response as CORRECT or INCORRECT according to the following guidelines.
> - INCORRECT responses deviate from the correct answer, use incorrect reasoning, are irrelevant, or do not directly answer the question.

---

[1] https://help.openai.com/en/articles/6825453-chatgpt-release-notes
[2] https://github.com/mmabrouk/chatgpt-wrapper

> - If a response gives correct reasoning but does not explicitly answer the question it should still be labeled as CORRECT.
>
> Note: Typos or grammatical errors should not be considered unless they affect the correctness of the answer.
> Here is the self-explanation question: `se_prompt`
> Here is the student's response: `student_answer`
> Here is the rubric for grading this question: `rubric_items`
> Please answer with only a single word: 'CORRECT' or 'INCORRECT'. Do not include any additional information.

**Fig. 2.** The query template for asking ChatGPT to grade a student's self-explanation answer. The placeholder fields `se_prompt`, `student_answer` and `rubric_items` are to be substituted for the actual data in each query.

For RQ3 – whether ChatGPT can provide instructionally meaningful feedback to incorrect self-explanation answers – we similarly constructed a query template, shown in Figure 3, which contains the self-explanation prompt (`se_prompt`), grading rubric items (`rubric_items`), and the student's answer (`student_answer`). After collecting ChatGPT's feedback responses for all 4,142 incorrect answers in the dataset, based on the human labeling from [21], we carried out a qualitative coding process [6] to evaluate them as follows. First, one researcher analyzed 1.5% of the responses to develop a coding rubric. Three members of the team then independently coded 20% of the responses based on this rubric, with an inter-rater reliability (IRR) score of at least 60% in each category. Next, the three members met to resolve disagreements in their codes and refine the rubric. Finally, they applied the updated rubric to code the remaining 80% of the data. The final rubric is included in Table 2.

> Suppose you are a middle school math teacher and you are currently teaching decimals. You want to grade students' explanations for how they solved a decimal question, as directed by a prepared question you wrote ahead of time. Given that question, a small rubric explaining the requirements for a correct answer, and the student's incorrect answer, please provide a clear and concise 1-3 sentences of feedback to the student about why their answer is incorrect and direct them to the correct answer.
> Here is the self-explanation question: `se_prompt`
> Here is the rubric for grading this question: `rubric_items`
> Here is the student's response: `student_answer`

**Fig. 3.** The query template for asking ChatGPT to provide feedback to a student's incorrect self-explanation. The placeholder fields `se_prompt`, `rubric_items` and `student_answer` are to be substituted for the actual data in each query.

**Table 2.** The final coding rubric for evaluating ChatGPT's feedback to incorrect answers.

| Category & Codes | Description | Cohen's κ |
|---|---|---|
| **Accuracy** (Yes, No, N/A) | Does ChatGPT accurately distinguish between partially correct and fully incorrect answers? Label N/A for questions with no partial credit. | 61.7% |
| **Fluency** (Yes, No) | Is the feedback written in grammatically correct and natural-sounding English? | 99.8% |
| **Regulation** (Yes, No, N/A) | Does the feedback address all decimal misconceptions reflected in the answer? Label N/A for answers that do not clearly reflect any misconception. | 87.3% |
| **Solution** (Yes, No, Incorrect) | Does the feedback tell the student the correct answer? | 69.9% |
| **Rationale** (Yes, No, Incorrect) | Does the feedback quote the appropriate rubric item(s) or provide a rationale to explain why an answer is incorrect? | 94.2% |
| **Encouragement** (Yes, No) | Does the feedback provide any form of encouragement? | 79.6% |

## 4 Results

### 4.1 Evaluation of ChatGPT's Decimal Skills

We observed that the ChatGPT interface provides the option to re-generate the response to a previous question, which leads to a different answer due to a change in the underlying random seed. To account for this variability, we had ChatGPT answer each of the 48 problem-solving questions and 24 self-explanation prompts in *Decimal Point* ten times and observed how consistently its answers were correct. Figure 4 shows ChatGPT's problem-solving and self-explanation performance in each problem type, as evaluated by a researcher in the team. Among the problem-solving questions, ChatGPT was able to correctly solve most of the *Bucket* and *Sorting* exercises, which both involve decimal number comparisons. However, it tended to claim that longer negative decimals were greater than shorter negative decimals (e.g., -1.701 > -1.7) and longer positive decimals were less than shorter positive decimals (e.g., 0.299 < 0.29), resulting in low correct rates at two *Bucket* problems. For decimal addition problems in the *Sequence* and *Addition* types, the majority of ChatGPT's numeric outputs were correct, but it did not provide the carry digits in *Addition* problems. Finally, ChatGPT struggled with *Number Line* problems the most, likely due to the need to visualize spatial distances that extends beyond text generation. Interestingly, in many cases ChatGPT could draw a valid number line and even correctly state where the given number should be located, but it still placed the number marker at an incorrect location.

With the self-explanation prompts, using the same rubric employed by [21] to grade student answers, we observed that ChatGPT was able to answer most questions correctly across all five problem types. There were only two questions in the *Sorting* type where ChatGPT provided incorrect reasoning. In particular, ChatGPT claimed that 9.222 is bigger than 9.2111 because "the digit 2 in the thousandths place of 9.222 is larger than the digit 1 in the hundredths place of 9.2111" and that 1.1 is bigger than 1.0111 "because both numbers have the digit 1 in the tenths place but the following digits are smaller in 1.0111 than in 1.1." These errors can be attributed to incorrect applications of decimal comparison by place values, and to mis-identifications of the values at each decimal place.

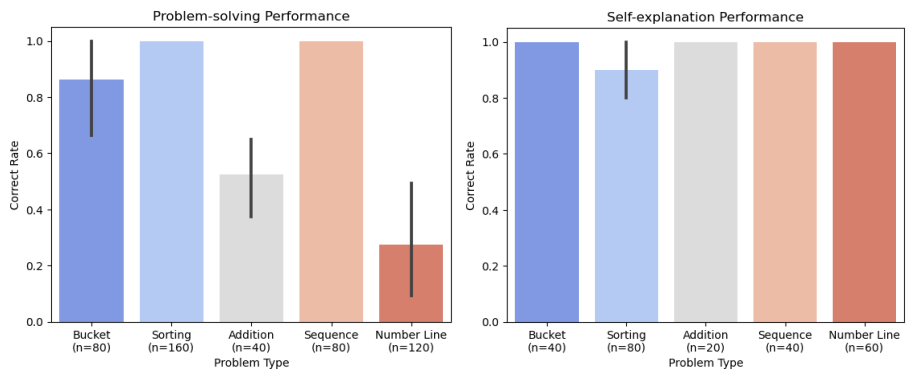

**Fig. 4.** ChatGPT's rate of correctness in the problem-solving and self-explanation questions, aggregated by problem type. Error bars denote the 95% confidence interval.

### 4.2 Evaluation of ChatGPT's Grading

Compared to the 1,000 correct and 4,142 incorrect labels of the self-explanation responses assigned by human graders in [21], ChatGPT identified 1,948 correct and 3,194 incorrect responses. Using the human labels as ground truth, we computed ChatGPT's accuracy as 0.75, ROC-AUC score as 0.78, and IRR as Cohen's $\kappa = 41.2\%$, with the full confusion matrix included in Table 3. We observed that most of the mismatches were due to ChatGPT grading an incorrect answer as correct, suggesting that it did not follow the grading rubrics as closely as human graders.

**Table 3.** The confusion matrix between humans' and ChatGPT's labels of the 5,142 self-explanation responses.

|  | Human: Correct | Human: Incorrect |
| --- | --- | --- |
| ChatGPT: Correct | 830 | 1,118 |
| ChatGPT: Incorrect | 170 | 3,024 |

The 1,118 cases of false positives are distributed across problem types as follows: 101 in *Bucket*, 174 in *Sorting*, 218 in *Addition*, 441 in *Sequence*, and 184 in *Number Line*. When analyzing the mismatches in each problem type, we observed the following patterns. For *Bucket* and *Sorting* self-explanation prompts, ChatGPT tended to focus on the presence of keywords such as "bigger" and "smaller" to determine correctness, regardless of whether the student provided an explanation or answered the question correctly (e.g., "-0.9 is smaller than 0.6 because it is smaller" was rated as correct by ChatGPT). In *Addition* problems, where the self-explanation prompts involve describing the process to line up two decimal numbers before adding them, ChatGPT tended to label student responses that included positional and directional words such as "underneath," "above," "left," and "right" (e.g. "underneath each other") as correct, even if they did not mention lining up by the decimal point or place values. In *Sequence* problems, students were asked to describe the general rule for finding the next numbers in a sequence, but many answers that were overly specific to the current mini-game (e.g. "to add 1.3 to whatever the last decimal is.") were regarded as correct by ChatGPT.

Finally, we observed from *Number Line* responses that the provision of an answer, without the accompanying reasoning, was sufficient to get a correct label from ChatGPT. In general, ChatGPT appeared to rely on particular keywords extracted from the rubric to determine correctness, rather than the content of the explanations.

### 4.3   Evaluation of ChatGPT's Feedback

The 4,142 feedback paragraphs generated by ChatGPT had an average length of 3.4 sentences (*SD* = 0.9) and contained 64.5 words on average (*SD* = 19.5). Figure 5 shows the evaluation results of these feedback messages, based on the categories included in Table 2. We observed that all of ChatGPT's outputs demonstrated proficient English usage, with a 100% rating of YES in the *Fluency* category. 716 of its feedback messages also contained some forms of encouragement, including "your answer is on the right track," "great job" and "keep practicing," but there were no clear patterns for when ChatGPT decided to provide encouragement. Additionally, ChatGPT was able to detect 9 cases of offensive or disrespectful language that students used in their self-explanations, likely due to their frustration, and reminded them to maintain an appropriate attitude in class.

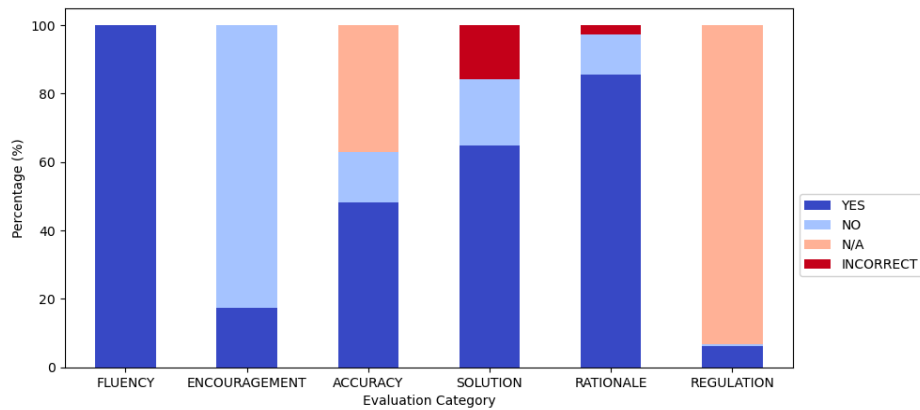

**Fig. 5.** The distribution of code labels for ChatGPT's feedback in each evaluation category.

For the *Accuracy* category, among the 2,607 incorrect answers to the self-explanation questions where partial credit is available (e.g., "Is 0.456 to the left or right of the number line? How do you know?"), ChatGPT was able to distinguish between partially correct answers and fully incorrect answers 1,994 times. Partially correct answers are those that provide a correct statement (e.g., "to the right") but incorrect or missing explanations, or vice versa. However, there were 110 cases where ChatGPT mistook a partially correct answer for being fully correct and simply congratulated the student, instead of explaining their mistakes. These answers typically lacked the explanation component required in the grading rubric, suggesting that ChatGPT was sometimes unable to apply all of the rubric criteria in its feedback generation.

Additional instances of ChatGPT not following the input query are observed in the *Solution* category, where it did not provide the solution in its feedback to 794 incorrect answers, even though it was explicitly requested to do so (Figure 3). In these cases, we observed that the students' answers were either empty, very brief or off-topic (e.g., "idk," "subtract," "by adding up"). In response, ChatGPT chose to remind the student of the relevant rubric items, instead of giving them the solution. However, there were also 659 cases where ChatGPT provided a solution that was incorrect. These cases predominantly fell into the *Sorting*, *Bucket* and *Number Line* problem types, where ChatGPT's solutions demonstrated similar issues as those in its responses to the problem-solving and self-explanation questions. In particular, with *Bucket* and *Sorting* problems, ChatGPT either mis-identified the decimal place values (e.g., "0.22 is bigger because it has a 2 in the tenths place value, whereas 0.2 only has a 0 in that place") or applied place values incorrectly in its comparisons ("9.222 is bigger than 9.2111 because the digit in the thousandths place is larger"). Likewise with *Number Line* problems, ChatGPT showed misconceptions about the number placement, even when its rationale was correct (e.g., "0.456 is a decimal number between 0 and 1, and since it is less than 0.5, it must be to the left of 0 on the number line").

While ChatGPT had occasional issues with generating its own solutions, it demonstrated good capabilities in understanding the students' answers and explaining the rationale for their incorrectness, as observed in 3,549 cases in the *Rationale* category. Our analysis showed that ChatGPT was able to provide a wide range of nuanced explanations to students, such as "your answer is not specific enough" or "good effort, but the reason why -8.517 is smaller than 8.5 is not explained in your answer." In many cases, it was able to directly reference the incorrect components of the students' answers. For instance, when responding to an incorrect answer, "[-0.07 is] closer to 0 because it has more digits," ChatGPT correctly pointed out that "the number of digits does not determine the closeness of a decimal number to 0 or -1." In general, while the majority of incorrect answers did not clearly reflect any misconceptions, due to students omitting the reasoning or not clearly expressing their thoughts, when misconceptions did manifest, ChatGPT was effective at addressing them in its feedback, as can be seen by the results of the *Regulation* category.

## 5   Discussion and Conclusion

In this work, we examined ChatGPT's capabilities as a "teaching assistant" to support student learning of decimal numbers in the digital learning game *Decimal Point*. Our goal was to derive actionable insights on the strengths and weaknesses of ChatGPT, and to identify its appropriate use cases in the classroom. Our first analysis of ChatGPT's responses to the problem-solving and self-explanation questions in the game revealed varying performance levels across decimal problem types. ChatGPT generally did well in decimal comparisons, but tended to make errors with numbers having three or more decimal digits. It could perform decimal additions and carrying correctly, but could not express when carrying took place. Finally, just as *Number Line* problems have been identified as the most difficult for students in prior studies [19], so did they pose the stiffest challenge to ChatGPT. Across the *Number Line* problems in the game, ChatGPT only answered one or two correctly out of ten, indicating a performance worse than chance. While identifying the cause of this poor performance can be difficult, as is typical with black-box neural networks, a feasible high-level explanation is that, as a probabilistic language model [2], ChatGPT is trained to output the most likely responses, rather than relying on internal mathematical rules. Thus, if a setting frequently appears in its training data, such as drawing a number line from -1 to 1, ChatGPT will be able to replicate it well, but will encounter problems with more specific tasks, such as locating 0.579 on the drawn number line. This line of reasoning also explains why ChatGPT had better performance in the self-explanation questions, where it could make better use of its language skills. We observed that ChatGPT was able to cite several textbook mathematical procedures almost verbatim (e.g., "if the sum of the digits in any column is greater than 9, you need to carry over the excess to the next column") to deliver satisfactory explanations, although it would still have occasional issues with applying these procedures to rarely occuring numbers such as 1.0111. Thus, at the moment, ChatGPT may be more suited for more conceptual rather than procedural math questions, and teachers should always double check its answers before sharing with their students.

Our second research question tests ChatGPT's comprehension skill, as we provide it with data from multiple sources – including the grading context, rubric items, input question and student's answer – and ask it to determine the correctness of the answer. Comparing ChatGPT's labels to ground-truth human labels from [21], we found that ChatGPT had low false negatives (170) but high false positives (1,118). Further analyses showed that these mismatches are likely due to shallow keyword matching with the grading rubric, rather than a rigorous application of the provided rubric items. Because this correctness classification task is fairly specific to the questions and answers in the learning game *Decimal Point*, it is possible that ChatGPT, as a general-purpose task solver [24], could not detect all of the nuances in the grading rubric. In this case, its performance can be improved by fine-tuning the underlying model (GPT-3.5) on a portion of the human-labeled data from [21] to acquire more accurate domain embeddings; however, doing so would take away ChatGPT's ease of use and require deep learning expertise to scale to other knowledge domains and learning contexts.

Next, we performed a thorough analysis of ChatGPT's feedback messages to all of the 4,142 incorrect self-explanations in the dataset. Our most interesting insight is that ChatGPT exhibited several behaviors that fit the "suppose you are a middle school math teacher" setting in our input query, even though some of these behaviors were not explicitly specified, or even contradicted our other requests. First, it demonstrated fluent English communication, provided various forms of encouragement, and also called out students' usage of inappropriate language. Second, although we had asked ChatGPT to provide the solution in its feedback, ChatGPT only reminded students about the rubric and prompted them to try again when it detected low-effort submissions, i.e., answers that were empty, too brief or off-topic. This behavior is consistent with prior research reporting that, instead of being shown the solution right away, students should go through a stage of productive struggle to attain deep learning [27]. Third, despite the frequent grammatical errors and misspellings made by students, ChatGPT demonstrated a good understanding of the students' answers and was frequently able to point out which components were incorrect, as well as to address the underlying misconceptions. The primary weakness of ChatGPT's feedback lies in its provision of the solutions, which exhibit several patterns of incorrectness, as discussed above.

Given the above insights, we propose that ChatGPT can assist teachers in the initial review and feedback generation stage, where it can point out prominent errors, misconceptions and suggest a general feedback outline. However, due to potential inaccuracies in its mathematical statements, ChatGPT is not yet ready to be used as a standalone student-facing application, and all of its outputs should be double checked by a domain expert. At the same time, several improvements to ChatGPT, including the GPT-4 model with superior mathematics performance [3], are already released or under active development. Thus, future research should closely monitor the progress of these LLMs to identify the best opportunity to integrate them in digital learning platforms.

Finally, we should note the limitations of this current work. First, we have only investigated ChatGPT's performance in the domain of decimal numbers. While other research has tested ChatGPT on a wider range of mathematical tasks [8], due to the complexities underlying LLMs, it is difficult to infer whether ChatGPT's good performance in one domain will transfer to another domain. Thus, researchers and instructors should perform their own assessments to determine the appropriate use cases of ChatGPT in their respective domains. Second, because ChatGPT's outputs depend on a set of random seeds which the user does not have access to, our current results may suffer from a lack of replicability. We expect that the development and release of future LLMs will be more streamlined and offer users more control over their interaction with such models. In the meantime, we have shared our scripts for querying ChatGPT on GitHub[3] to support the research community in reproducing our data collection process. We envision this work as a foundational step towards more rigorous benchmarks for assessing the roles of LLMs in education. Ultimately, the field of LLMs would greatly benefit from the inputs of both AI researchers and educators alike to ensure its contribution towards scalable, equitable and effective technology-enhanced learning.

**Acknowledgements.** This work was supported by NSF Award #DRL-2201796. Thanks to Injila Adil for assisting with the coding of the self-explanation feedback.

---

[3] https://github.com/McLearn-Lab/ECTEL2023-DP-SE